# A Computational Model of YAP/TAZ Mechanosensing

**Keywords**: pathway model, Hippo pathway, synergy, adhesion, cytoskeleton dynamics

Meng Sun[1], Fabian Spill[2,1,*] and Muhammad H. Zaman[1,3,*]

1. Department of Biomedical Engineering, 44 Cummington Street, Boston MA 02215, USA
2. Department of Mechanical Engineering, Massachusetts Institute of Technology, 77 Massachusetts Avenue, Cambridge, MA 02139, USA
3. Howard Hughes Medical Institute, Boston University, Boston, MA 02215, USA

*Corresponding author

E-mail: zaman@bu.edu (MHZ) and fspill@mit.edu (FS)

# Abstract

In cell proliferation, stem cell differentiation, chemoresistance and tissue organization, the ubiquitous role of YAP/TAZ continues to impact our fundamental understanding in numerous physiological and disease systems. YAP/TAZ is an important signaling nexus integrating diverse mechanical and biochemical signals, such as ECM stiffness, adhesion ligand density, or cell-cell contacts, and thus strongly influences cell fate. Recent studies show that YAP/TAZ mechanical sensing is dependent on RhoA-regulated stress fibers. However, current understanding of YAP/TAZ still remains limited due to the unknown interaction between the canonical Hippo pathway and cell tension. Furthermore, the multi-scale relationship connecting adhesion signaling to YAP/TAZ activity through cytoskeleton dynamics remains poorly understood. To identify the roles of key signaling molecules in mechanical signal sensing and transduction, we present a novel computational model of the YAP/TAZ signaling pathway. This model converts ECM mechanical properties to biochemical signals via adhesion, and integrates intracellular signaling cascades associated with cytoskeleton dynamics. We perform perturbations of molecular levels and sensitivity analyses to predict how various signaling molecules affect YAP/TAZ activity. Adhesion molecules, such as FAK, are predicted to rescue YAP/TAZ activity in soft environments via the RhoA pathway. We also found that changes of molecule concentrations result in different pattern of YAP/TAZ stiffness response. We also investigate the sensitivity of YAP/TAZ activity to ECM stiffness, and compare with that of SRF/MAL, which is another important regulator of differentiation. In addition, the model shows that the unresolved synergistic effect of YAP/TAZ activity between the mechanosensing and the Hippo pathways can be explained by the interaction of LIMK and LATS. Overall, our model provides a novel platform for studying YAP/TAZ activity in the context of integrating different signaling pathways. This platform can be used to gain new fundamental insights into roles of key molecular and mechanical regulators on development, tissue engineering or tumor progression.

# Introduction

One of the fundamental questions of cell biology is how cells organize themselves into complex tissues and three-dimensional structures. Multiple signals transmitted from neighboring cells and extracellular matrix (ECM) environment are integrated into intracellular pathways and guide cellular behavior, such as proliferation, migration and differentiation. Recent years have seen enormous progress in uncovering the roles of the transcriptional regulator YAP/TAZ, which controls organ size by integrating both physical and biochemical cues (1–3). First discovered in *Drosophila*, deregulation of YAP or its paralog TAZ has been found to lead to massive organ overgrowth (4). YAP/TAZ has been shown to have significant roles in regulating remarkable biological properties in development, tissue homeostasis and cancer. The activation of the most studied upstream cascade of YAP/TAZ, the canonical Hippo pathway, is mainly dependent on cell density sensing, and results in the activation of the core Hippo kinase LATS which regulates YAP/TAZ localization (1). Despite these ubiquitous roles, the mechanism of YAP/TAZ localization regulation is still poorly understood (2). One of the models suggests that activated LATS in the Hippo core complex prevents YAP/TAZ from entering the nucleus via phosphorylation, involved with further cytosolic sequestration through binding with molecules such as 14-3-3, or degradation (2, 5). YAP/TAZ lacks of a nuclear localization signal (NLS), and the machinery for their nuclear import/export is unknown (2). Thus it is an important question to investigate which molecules are involved in this nuclear transport regulation, such as RanGTPases (6). After translocating into the nucleus, YAP/TAZ acts as a transcriptional co-activator with several transcription factors and thus targets various genes, including TEAD (2). More recently, attention has been focused on mechanical regulators of YAP/TAZ activity beyond the core Hippo pathway (7). For instance, increasing stiffness of the external environment (8), or exerting static stretch (7) has been found to promote YAP/TAZ nuclear translocation and its downstream transcription activity. Specifically, it was found that these effects are transmitted through RhoA pathways and resulting stress fiber formation (8). Therefore, YAP/TAZ can be considered as master-integrators of the cellular microenvironment, integrating different cues of mechanical and biochemical nature.

Sensing mechanical properties of ECM alone is a complex process and of fundamental importance in regulating cell behaviors (9, 10). Stiff or high adhesion environments have been shown to promote both migration and proliferation (11). For instance, cells have been shown to migrate towards a stiffer ECM environment (durotaxis) (12) or a surface-bound gradient of ECM adhesion (haptotaxis) (13). Furthermore, manipulation of ECM stiffness can direct mesenchymal stem cells to differentiate into neuron, myoblasts or osteoblasts in a myosin II-dependent manner (14). Up-regulating YAP activity on soft substrates is shown to direct mesenchymal stem cell differentiation towards osteogenesis, which otherwise is observed on stiff substrates (8, 15). Specifically, the sensing of the mechanical properties of the ECM relies on force-sensitive molecules located at cell-ECM adhesion complexes, converting the mechanical signals to biochemical signals inside the cell (16, 17). These adhesion complexes form around integrins, which are membrane-bound heterodimers, and bind to the ECM. Important downstream targets include FAK (18, 19), which is phosphorylated at tyrosine 397 (20) upon integrin binding, and Src further enhances FAK phosphorylation sites at other residues to maximize the kinase activity of FAK (19, 21). Other than adhesion sensing, phosphorylated FAK at tyrosine 397 is also sensitive to the stiffness of the surrounding matrix in both 2D and 3D environments (11, 22, 23). Importantly, in 3D, FAK diffuses throughout the cytoplasm but still modulates cell motility by affecting protrusion activity and matrix deformation (24). 2D substrates have demonstrated that cells increase their assembly of F-actin stress fibers and focal adhesions via RhoA contractility regulation when substrate rigidity is increased (9, 13, 25). Stiffness sensing in 3D is also shown to be associated with RhoA dependent cytoskeleton regulation (23, 26).

Mechanical sensing of the microenvironment is a complex and multi-step process with profound implications for maintaining homeostasis and disease relevance (9, 10, 27, 28). Various quantitative aspects of signal transmission from the ECM stiffness or adhesion to intracellular biochemical signals remain largely elusive. While there are several mechanical models (29), and multiple studies of cytoskeleton dynamics, there is a lack of an integrated model of mechanosensing and YAP/TAZ which enables one to understand the interactions of various factors influencing YAP/TAZ activity in a quantitative way. Furthermore, such integrated model can be used to predict the effect of molecular inhibitors or changes in mechanical properties *in silico*. In this paper, we aim to bridge this gap through an integrated model that includes mechanical-to-biochemical signal conversion by adhesion molecules, intracellular signal transmission, cytoskeleton dynamics, and regulation of effectors relevant to directing transcriptional programs, such as YAP/TAZ (30). We introduce a response function for YAP/TAZ activity dependent on ECM stiffness, and study how disturbances of molecular activities can alter this stiffness response function. The model predicts that adhesion regulation has a significant role in regulating YAP/TAZ activity in mechanical sensing. Changes in FAK shift the stiffness response function horizontally, such that FAK overexpression rescues YAP/TAZ activity in soft environments, while mDia shifts the stiffness response function vertically. We also predict that YAP/TAZ is more sensitive than SRF/MAL in response to changes in ECM properties. Furthermore, our model is able to explain the observed synergistic effect of mechanical and canonical Hippo regulation on YAP/TAZ activity through incorporating interaction between LATS and LIMK.

# Methods

To understand and predict YAP/TAZ activity, as measured by its amount in the nucleus, we consider a model of interactions among several key molecules involved in mechanical sensing mechanism (14, 31). Previous models focused on providing new insights into individual components of mechanical signaling, such as adhesion (32), Rho GTPase (33), and cytoskeleton dynamics (34–37) separately. However, an integrated understanding of how adhesion affects cytoskeleton dynamics and stiffness response effector such as YAP/TAZ is not well characterized. Our model includes key selective molecules from previous models, such as FAK and RhoA, and is extended with the necessary components for YAP/TAZ regulation (8). We model the time evolution of the concentrations of the molecules per cell by ordinary differential equations. The model distinguishes between molecules in different cellular compartments, such as YAP/TAZ in the cytoplasm and nucleus, and F-actin in the cortex and the cytoplasm. It consists of the conversion of mechanical signals to biochemical signals via adhesion molecules, RhoA GTPase signal transmission, cytoskeleton dynamics, and the regulation of transcriptional activity (16, 30, 38).

*Signaling network*

The network of signaling molecules is shown in Figure 1. External ECM properties are felt through adhesion complexes via integrins. Higher stiffness of ECM correlates with larger protein aggregation at focal adhesions (39). Thus the clustering of integrins attracts more FAK to activate them (40). For this paper we focus on FAK, for which there is clear evidence of higher activity in stiffer 3D ECM environments (11, 41). Further downstream, we include the small GTPase RhoA, which can activate autoinhibited mDia (42, 43) and ROCK (44) by binding with them. mDia is an actin formin that nucleates actin filaments (45) and accelerates the rate of actin elongation 5- to 15- fold (46). ROCK enhances myosin activity by phosphorylating myosin light chain and inhibiting myosin phosphatase (47). It also phosphorylates and activates LIM-kinase, resulting in the inactivation of the F-actin severing protein cofilin (44). Thus ROCK promotes the assembly of stress fibers and actomyosin contractility. In turn, increased actin and myosin activity is associated with increased YAP/TAZ activity (7, 8). In our model these key molecules are sufficient to interpret the mechanism of mechano-sensing of YAP/TAZ, and the model is formulated in a way that other molecules, such as Rac, Cdc42, profilin, Aip1 (48), Src or capping proteins, can be easily added to the model. The inputs are the ECM mechanical properties and concentrations of intracellular signaling molecules, while the output is the nuclear translocation of the relevant transcriptional molecule, YAP/TAZ. The signaling process of another key regulator of differentiation, SRF/MAL (49), is used in comparison with YAP/TAZ mechanical sensing. The entry of MAL into the nucleus to interact with SRF transcription factor activity is regulated through monomeric G-actin. G-actin binds to MAL to prevents it from binding to SRF and thus its downstream transcription (50).

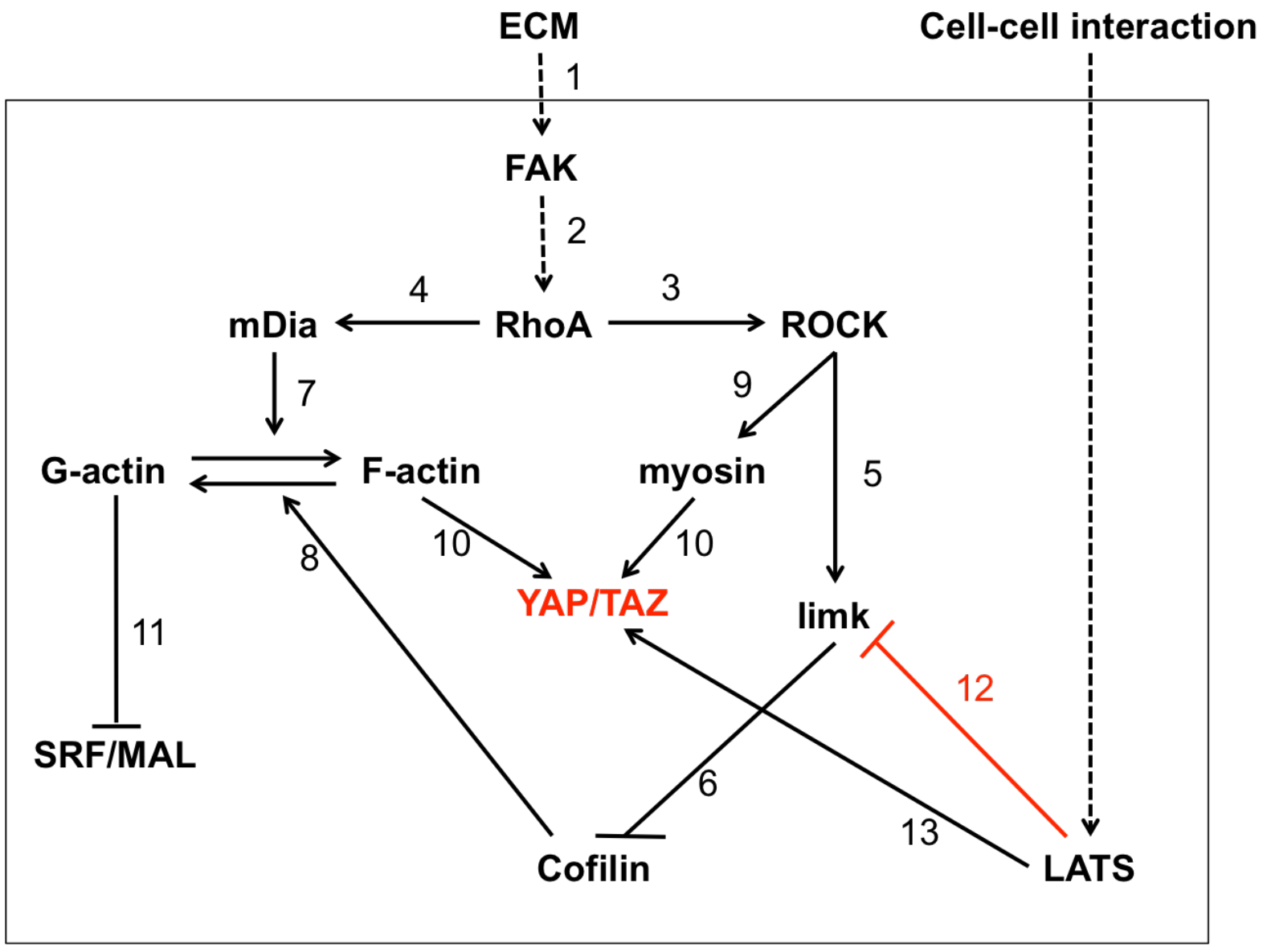

**Figure 1 YAP/TAZ mechano-sensing signaling pathway network**

This is the scheme of the computational signaling cascade model. The ECM mechanical properties are transmitted to intracellular signals via adhesion molecules, such as FAK. Adhesion molecules activation induces RhoA binding with GTP. The RhoA-GTP drives the formation of actomyosin via mDia and ROCK activation. mDia is a formin that polymerizes F-actin. ROCK can both activate myosin and induce F-actin assembly through LIMK and cofilin. Eventually, the resultant of F-actin and myosin activity, i.e., stress fibers in 2D systems, leads to the YAP/TAZ nuclear translocation. In comparison with the YAP/TAZ mechanical sensing mechanism, SRF/MAL senses mechanical inputs by being inhibited by G-actin. In addition, the Hippo core component LATS is sensitive to cell-cell interaction. In our model, the synergistic effect between the mechanical sensing and the Hippo pathway is predicted due to the interaction of LATS and LIMK.

*Conversion from mechanical signals to biochemical signals*

ECM stiffness and ligand density are two key ECM properties initially sensed and transmitted to intracellular signals via adhesion molecules (16). Elucidating the mechanisms of conversion of mechanical signals to biochemical signals via adhesion molecules is a fundamental question of mechanobiology, and the precise mechanisms of mechanical sensing in regulating YAP/TAZ activity are poorly understood. Among the large number of adhesion molecules, we focus on the activity of FAK in our model, since FAK is a key molecule in adhesion components regulating the downstream signals, such as RhoA GTPase (51). Other components of adhesion molecules are implicitly incorporated into the model by the usage of a higher-order Hill function, which represents cooperativity among the groups of adhesion molecules. There is strong evidence showing that with higher rigidity of ECM, FAK phosphorylation is up-regulated both on 2D substrate and in 3D matrix (11, 22, 23, 41). Over-expression of FAK can rescue cells whose proliferation was inhibited by low adhesion or reduced cell spreading (52). In 2D, the cells constrained in spreading showed lower FAK phosphorylation than the spread cells, even when coated with high densities of fibronectin, and had impaired proliferative ability (52).

Previous modeling approaches have focused on the dependence of the FAK phosphorylation rate on the ECM ligand density through a Michaelis-Menten function (32, 53). We extend this function by adding the mechanical properties of ECM, notably its Youngs modulus $E_{mol}$. When fitting data regarding the dependence of FAK on ECM stiffness (11), we found that a second-order Hill function is able to capture the features of the data (Figure S1 in the Supporting Materials). This is consistent with previous studies showing that a Hill function is compatible with the high cooperativity of the adhesion molecules such as integrin, talin, etc (51). Hence, the time dependence of active phosphorylated FAK is given by

$$\frac{dFAK_P(t)}{dt} = k_{sf} \frac{(LD(t)\times E_{mol}(t))^2}{C^2 + (LD(t)\times E_{mol}(t))^2}(FAK^0 - FAK_P(t)) - k_{df} FAK_P(t) . \qquad (1)$$

Here, $k_{sf}$ is the maximum activation rate due to the ECM, and the de-phosphorylation rate of FAK is $k_{df}$. $C$ is the value of $LD(t)\times E_{mol}(t)$ when the activation rate of FAK is $k_{sf}/2$, that is, half of the maximum activation rate. We mainly focus on the ECM stiffness in this paper and regard ligand density as constant.

*Molecular switch RhoA GTPase*

Rho-family GTPases are molecular ‘switches’ within cells to control the formation and disassembly of actin cytoskeletal structures, such as stress fibers and filopodia. Adhesion molecules, such as FAK, sense the ECM and activate Rho-family GTPases. Among them, RhoA is a key GTPase which has been shown to be up-regulated with increasing rigidity of ECM in 3D environments (11, 41). In detail, phosphorylated FAK activates RhoA in both a direct (54) and indirect manner, such as via Fyn (55). Hence, we model the time-dependence of the concentration of RhoA as

$$\frac{dRhoA_{GTP}(t)}{dt} = k_{fk\rho}(\gamma (FAK_P(t))^2 + 1)(RhoA_0 - RhoA_{GTP}(t)) - k_{d\rho} RhoA_{GTP}(t) . \qquad (2)$$

Here, $k_{fk\rho}$ is the baseline activation rate of RhoA due to the molecules other than phosphorylated FAK, $k_{d\rho}$

is the RhoA-GTP deactivation rate, which is relevant to the hydrolysis rate of RhoA bound GTP to GDP, and $\gamma$ is the RhoA activation enhancement/amplification due to active FAK. We have modeled the dependence of the activation rate of RhoA on phosphorylated FAK as second order, so the steady state

$$RhoA_{GTP} = \frac{k_{fk\rho}(\gamma(FAK_p(t))^2+1)}{k_{d\rho}+k_{fk\rho}(\gamma(FAK_p(t))^2+1)}$$

is a quasi-second-order hill function. This can represent the cooperative process of activating RhoA through FAK as discussed earlier.

*Cytoskeleton regulation*

Downstream of RhoA, several cytoskeleton regulators are activated. We focus on ROCK and mDia to model their time-dependence by

$$\frac{dROCK_A(t)}{dt} = k_{r\rho}RhoA_{GTP}(t)(ROCK_0 - ROCK_A(t)) - k_{drock}ROCK_A(t), \quad (3)$$

$$\frac{dmDia_A(t)}{dt} = k_{m\rho}RhoA_{GTP}(t)(mDia_0 - mDia_A(t)) - k_{dmDia}mDia_A(t). \quad (4)$$

Here $k_{r\rho}$ is the activation rate of ROCK due to RhoA-GTP, $k_{drock}$ is the deactivation rate of ROCK, $k_{m\rho}$ is the activation rate of mDia due to RhoA-GTP, and $k_{dmDia}$ is the deactivation rate of mDia. Myosin is then up-regulated by ROCK, and evolves according to

$$\frac{dMyo_A(t)}{dt} = k_{mr}(\varepsilon T(ROCK_A(t))+1)(Myo_0 - Myo_A(t)) - k_{dmy}Myo_A(t). \quad (5)$$

Here $k_{mr}$ is the activation rate of myosin due to pathways other than ROCK, $k_{dmy}$ is the deactivation rate of myosin, and $\varepsilon$ is the myosin activation enhancement/amplification due to active ROCK.

Activated ROCK further activates LIM-kinase (LIMK) and LIMK phosphorylates cofilin to prevent its binding with actin and deactivates its severing and disassembly of F-actin.

$$\frac{dLIMK_A(t)}{dt} = k_{lr}(\tau T(ROCK_A(t))+1)(LIMK_0 - LIMK_A(t)) - k_{dl}LIMK_A(t), \quad (6)$$

$$\frac{dCofilin_{NP}(t)}{dt} = k_{turn-over}(Cofilin_0 - Cofilin_{NP}(t)) - k_{cr}(1-k_{ll}LATS_0)LIMK_A^2(t)Cofilin_{NP}(t). \quad (7)$$

Here, $k_{lr}$ is the activation rate of LIMK due to pathways other than ROCK, such as PAK (56), $k_{dl}$ is the deactivation rate of LIMK, $\tau$ is LIMK activation enhancement/amplification due to active ROCK compared with the activation rate due to other molecules, $k_{turn-over}$ is the dephosphorylation rate of cofilin, $k_{cr}$ is phosphorylation rate of cofilin due to LIMK, $k_{ll}$ is the inhibition rate of LIMK-dependent cofilin phosphorylation due to the total concentration of $LATS_0$ in the cell, and the function $T(ROCK_A(t))$ is a smoothing function given by

$$T(ROCK_A(t)) = smoothing\left( \begin{array}{c} 0, ROCK_A(t) \le ROCK_B \\ ROCK_A(t) - ROCK_B, ROCK_A(t) > ROCK_B \end{array} \right).$$

This T-function approximates the activation starting region of a Michaelis-Menten function. It mimics that active ROCK concentrations are required to exceed a threshold to activate LIMK. This approximates the flat region of the deactivating part of the Michaelis-Menten function. Once active ROCK is above the threshold concentration $ROCK_B$, the LIMK activation rate is linearly dependent on $ROCK_A(t)$, approximating the linear activation part of the Michaelis-Menten function. The smoothing function uses a smooth filter for connection, so the T-function and its first derivative are continuous. The saturation part of the Michaelis-Menten function is implicit in the T-function due to the saturation of concentrations of other molecules, such as FAK and RhoA.

Here we also propose a distinction between cortical F-actin and F-actin mainly in the interior cytoplasm, named as cytoplasmic F-actin, which is a key component forming stress fibers. The cortical F-actin and cytoplasmic F-actin have distinct regulators and downstream effectors. In *Drosophila*, *Tsr*, which is a cofilin homolog and accumulates F-actin around the entire cell cortex, does not have an effect in *Yki*, the YAP homolog, while the inhibition of *capt*, another homolog of cofilin, which restricts actin filament accumulation near the apical surface, promotes *Yki* activities (57). Cortical actin associating with apical junctions is also latrunculin resistant (58). Thus, in our model, we distinguish the cortical F-actin and cytoplasmic F-actin. For simplicity, we assume that the ratio between G-actin and cortical F-actin is constant. The cytoplasmic F-actin is regulated by mDia and cofilin as follows:

$$\frac{dFcyto(t)}{dt} = k_{ra}(aT(mDia_A(t)) + 1)Gactin(t) - k_{dep}Fcyto(t) - k_{fc1}cofilin_{NP}(t)Fcyto(t). \quad (8)$$

Here, $k_{ra}$ is the polymerization rate of cytoplasmic F-actin from G-actin, $k_{dep}$ is the depolymerization rate of cytoplasmic F-actin, $\alpha$ is the amplification/enhancement of the polymerization rate due to mDia, and $k_{fc1}$ is the severing and disassembly rate of cytoplasmic F-actin due to active cofilin. Cofilin is an F-actin assembly regulator and its effect is dependent on its concentration *in vitro*: if the ratio of cofilin/actin subunits in a filament is low (<1/100), this results in persistent filament severing and acceleration of F-actin depolymerization, while if the ratio is above 1/10, cofilin stabilizes F-actin in a twisted conformation and can nucleate filaments (56). The severing effect of cofilin itself can be biphasic when interacting with other cytoskeleton regulators such as Arp2/3 (35). Therefore in our model, cofilin's effect is to reduce the cytoplasmic F-actin concentration in the range of given endogenous cofilin and actin amounts (36). For simplicity, we do not explicitly take cortical actin into account and assume it is proportional to G-actin with a ratio $\beta$.

*YAP/TAZ dynamics*

YAP/TAZ is regulated by both stress fiber/RhoA activity and cell-cell contacts/LATS activity (2). It translocates to the nucleus to bind with transcription factors such as TEAD and activate them. Phosphorylation by LATS or, potentially through some currently unknown kinase whose activity is regulated by cytoskeletal tension, leads to sequestration of YAP/TAZ. We thus model nuclear YAP/TAZ concentrations by

$$\frac{dYAP/TAZ_N(t)}{dt} = (k_{CN} + k_{CY} Fcyto^n(t) myo_A(t))(YAP/TAZ_0 - YAP/TAZ_N(t)) - k_{NC} YAP/TAZ_N(t) - k_{ly} LATS_P(t) YAP/TAZ_N(t). \quad (9)$$

Here, $k_{CN}$ is the rate of YAP/TAZ trans-locating from the cytoplasm to the nucleus with no active cytoplasmic F-actin and myosin, $k_{NC}$ is the rate of YAP/TAZ translocating from nucleus to cytoplasm with no active LATS, and $k_{CY}$ is the YAP/TAZ nuclear translocation rate due to the stress fibers or the tensional cytoplasmic F-actin. $n$ characterizes the weight of how YAP/TAZ activity is dependent on cytoplasmic F-actin. $n = 1$ is the default setting that indicates the myosin and cytoplasmic F-actin have a similar effect on regulating YAP/TAZ activity, and $k_{ly}$ is the cytoplasmic translocation rate of YAP/TAZ due to active LATS. The mechanism for YAP/TAZ nuclear import/export is unknown (2), thus we model the YAP/TAZ nuclear transport directly depending on the stress fibers and LATS as approximation. Importantly, YAP/TAZ nuclear translocation has been shown to dependent on stress fibers and RhoA in 2D (8). In 3D stress fibers in cells are not as distinct as those present in the 2D monolayer cultures (59, 60), but still with more F-actin in the cytoplasm in rigid environments in comparison to the softer ones (61). Thus for cells with no apparent stress fiber structure in 3D, we generally use the tensional cytoplasmic F-actin filaments, which are characterized by the product of the cytoplasmic F-actin concentration and active myosin concentration $Fcyto(t) myo_A(t)$, to be an equivalent component of stress fibers in 3D.

*MAL dynamics*

MAL in the cytoplasm has been shown to bind with G-actin, which inhibits its ability to go to the nucleus and associate with SRF to form the activate SRF/MAL complex (50).

$$\frac{dMAL_N(t)}{dt} = \frac{k_{cnm}}{1 + k_{mg} Gactin(t)^2}(MAL_0 - MAL_N(t)) - k_{ncm} MAL_N(t). \quad (10)$$

Here, $k_{cnm}$ is the rate of MAL translocation from the cytoplasm to the nucleus, $k_{mg}$ is the decreasing effect due to cytoplasmic MAL binding with G-actin and retained in the cytoplasm (this inhibition effect is a second-order hill function), and $k_{ncm}$ is the rate of MAL trans-locating from the nucleus to the cytoplasm.

# Results

We use our computational model to investigate how mechanical signal transmission and YAP/TAZ regulation depend on pathways of interacting signaling molecules. We study the effect of kinetic inhibition, and perform local sensitivity analysis to predict the role of crucial model parameters and external variables such as stiffness. The model not only reproduces the existing experimental results, but also makes a number of new predictions. First, we study the effect of disturbing adhesion complexes on YAP/TAZ activity represented by the stiffness response function. Then, we explore the distinction between SRF/MAL and YAP/TAZ sensitivity to ECM stiffness. More importantly, we investigated the unresolved synergistic effect between the mechanical sensing and the Hippo pathway of YAP/TAZ through the addition of interaction between LIMK and LATS.

*Molecular intervention*

To demonstrate the fidelity of our model, we reproduced the results observed in experiments with molecular intervention, so that the model can be readily used to make novel predictions. The YAP/TAZ signaling cascade in response to mechanical stimuli is studied mostly by molecular interventions targeting the actin cytoskeleton, such as by adding blebbistatin or latrunculin A (8). We mimicked the inhibition or overexpression process in our model by upregulating or downregulating the corresponding kinetic parameters. The results are consistent with the experimental ones from Dupont et al (8) (Figure 2a). Furthermore, we examined the sensitivity of YAP/TAZ activity to local changes in kinetic parameters (62). It turns out that the parameters relevant to cell-ECM adhesion are the ones most robustly affecting YAP/TAZ activity (Figure 2b and S2 in the Supporting Materials), so we study their effect in depth in the following sections.

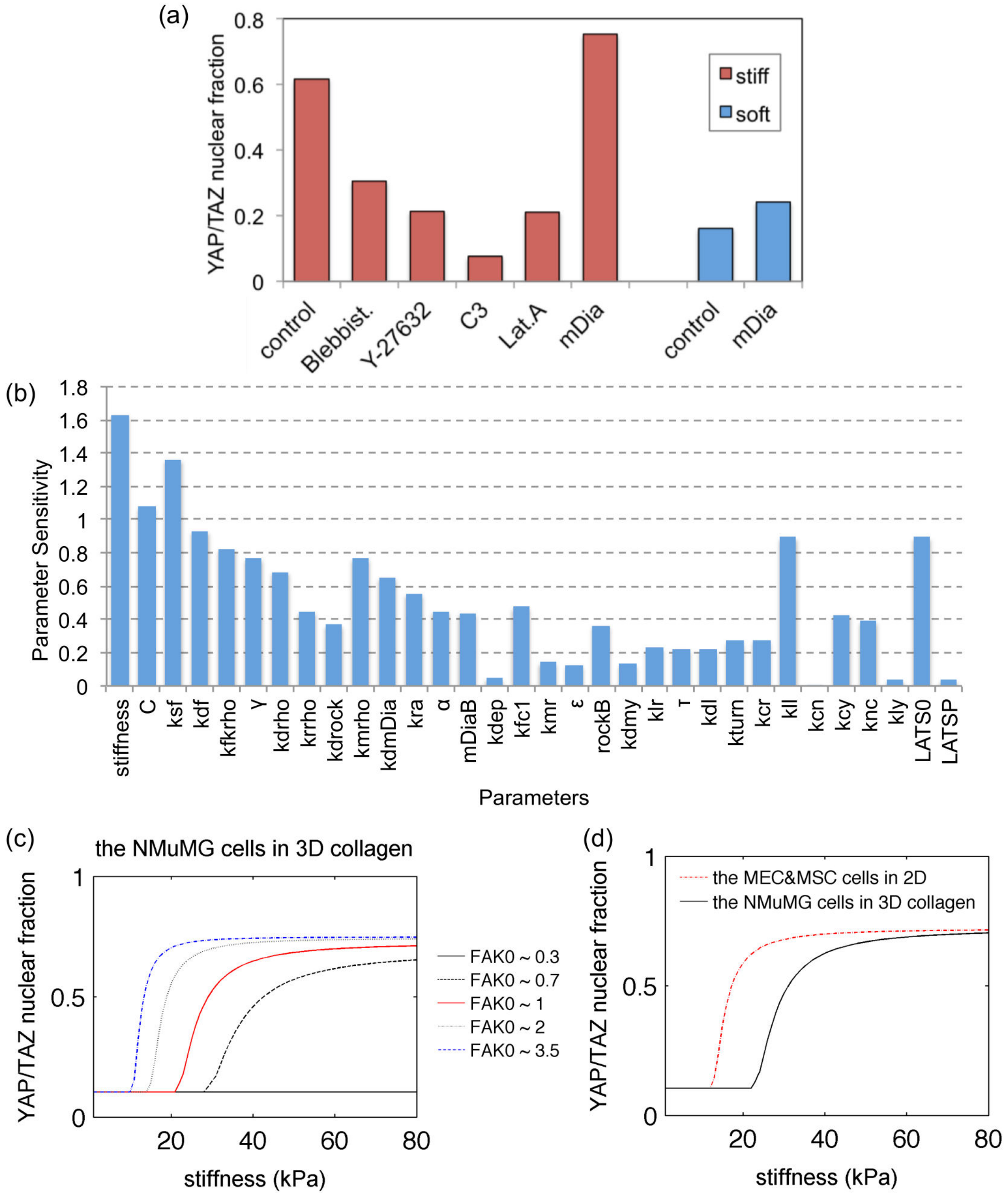

**Figure 2 Molecular intervention of YAP/TAZ activity in stiffness sensing**

(a) The results in the computational model regarding molecular interventions are consistent with the experimental results from previous mechanical regulation studies on YAP/TAZ (8). Here the molecular intervention process is mimicked by upregulating or downregulating the corresponding kinetic parameters (Supporting Methods in the Supporting Materials). The soft environment here is 0.5kPa hydrogel for 2D MEC cells, while stiff environment is 20kPa hydrogel(8). Blebbist.: myosin inhibitor. Y-27632: ROCK inhibitor. C3: RhoA inhibitor. Lat. A: F-actin inhibitor. mDia: over-expression of mDia (a formin that induces F-actin assembly). (b) Local sensitivity analysis of the nondimentionalized model. The local perturbations of adhesion-relevant parameters ($C$ , $k_{sf}$ and $k_{df}$ ) are the most robust ones in affecting YAP/TAZ activity in the model. This is the local sensitivity analysis of YAP/TAZ activity in stiff environment by downregulating the kinetic parameters individually 10%. The upregulating effect is much weaker than this, since most of the molecular activities are saturated. The up-regulation in a soft environment also shows that the adhesion is the most robust component in affecting YAP/TAZ activity (Figure S2 in the Supporting Materials). (c) and (d) are the cell stiffness response function, which is a YAP/TAZ activity function regulating cellular behavior such as proliferation dependent on external ECM stiffness. (c) Different amounts of FAK determine the threshold stiffness for cell stiffness sensing by shifting the curves horizontally. FAK0~1 corresponds to the endogenous FAK amount of NMuMG cell in 3D collagen. Collagen with 25 kPa is a soft environment and 45 kPa is a stiff environment for 3D NMuMG cells. If FAK is over-expressed to about 3-fold, 20 kPa will appear as a stiff environment for these cells. (d) The 2D MEC and MSC stiffness sensing identifies 1kPa ECM environment as soft and 20kPa as stiff, while in 3D NMuMG cells senses 25kPa as soft and 45kPa as stiff. These stiffness response functions are very alike the ones of manipulating only the FAK total amount of cells in (c).

The initiation of YAP/TAZ signaling converts the biomechanical signals into molecular signals. The first stage in the cascade, as described in the Methods section and Figure 1, is the phosphorylation level of FAK in response to ECM stiffness. Overexpressing or downregulating FAK amount has been demonstrated to be able to rescue or inhibit cell proliferation (52), respectively. Furthermore, overexpression of FAK has shown to rescue YAP/TAZ activity in serum-starved cells in a PI3K- and LATS-dependent way (63). However, with sufficient EGF in the environment, we predict that FAK overexpression can rescue YAP/TAZ activity in a soft ECM environment dependent on the RhoA cascade (Figure 2c). For example, with undisturbed FAK levels, the cells show high YAP/TAZ activity in ECM with a high stiffness of 45kPa, while they have low YAP/TAZ activity with 25kPa ECM stiffness. On the other hand, if $FAK_0$ is overexpressed by severalfold, YAP/TAZ activity is rescued in ECM environment as soft as 20 kPa. More specifically, FAK-overexpression rescues YAP/TAZ by shifting the stiffness response function leftwards rather than upwards. The transition window in ECM stiffness between high and low YAP/TAZ activity is also shortened with FAK-overexpression. Thus, we further predict that in downregulated FAK cells, activating YAP/TAZ activity requires considerably higher stiffness than in cells with normal FAK levels. That is, the YAP/TAZ activity in downregulated FAK cells can be rescued by plating them in a much stiffer environment. Changing mDia concentration mainly shifts the YAP/TAZ stiffness response functions vertically (Figure S3 in the Supporting Materials), and so do many actin regulators. In addition, increasing FAK concentration results in a narrower range of rigidity sensing and response in 3D, which displays high sensitivity of YAP/TAZ activity to stiffness between 10 and 15 kPa but little sensitivity in the range of 25-40 kPa.

In addition, comparing the cell stiffness response functions in 2D and 3D environments (Figure 2d), we found that their difference in YAP/TAZ activity is very similar to the difference of stiffness response functions with changing concentrations of FAK (Figure 2c). For example, for normal murine mammary gland cells (NMuMG) in 3D environments, 45 kPa appears as a stiff environment in terms of YAP/TAZ activity, whereas 25 kPa appears as soft. On the other hand, for human mammary epithelial cells (MEC), 20kPa appears as stiff whereas 1kPa appears soft. In agreement with the stiffness response function due to FAK concentration change, FAK has been shown to be more aggregated around the cell membrane in a 2D environment and thus has a higher total concentration than in a 3D environment, in which FAK is diffused throughout the cytoplasm (24) and has a lower average concentration. Thus, different localization of FAK in 2D and 3D environments may explain differences of stiffness sensing and cell behavior in such environments. Due to differences for varying cell types and dimensions, in the following we talk about the cell behavior and stiffness environment generally, rather than clarifying specific cell types, dimensions or ECM stiffness values.

*Comparison between YAP/TAZ and SRF/MAL*

In a single-cell mechanical regulation of YAP/TAZ activity is thought to be transmitted mainly though stress fibers (8), which are a resultant of both F-actin assembly and myosin activity, while SRF/MAL nuclear translocalization is mainly regulated by the amount of G-actin in the cytoplasm (49, 50). One of the hypotheses presented in the literature is that due to the resultant interaction of F-actin and myosin, YAP/TAZ nuclear localization is more sensitive to the stiffness of the ECM than SRF/MAL, which is only affected by actin. To better study the effect of each molecular perturbation on the YAP/TAZ nuclear translocation in comparison with MAL nuclear translocation, we define the fold-change of nuclear YAP/TAZ fraction when comparing soft and stiff ECM environment,

$$FoldChange = \frac{(YAP/TAZ_N)_{stiff} - (YAP/TAZ_N)_{soft}}{(YAP/TAZ_N)_{soft}}. \qquad (11)$$

This fold-change characterizes the sensitivity of YAP/TAZ activity with regard to stiffness, that is, soft versus stiff. SRF/MAL sensitivity to the ECM environment is also defined in the same way as Eq. 11. YAP/TAZ activity was found to be more sensitive to ECM stiffness than SRF/MAL signaling (Figure 3b and S4 in the Supporting Materials). SRF/MAL is clearly sensitive to F/G-actin, while there is no clear evidence of SRF/MAL being a genuine mechano-sensor. In addition, YAP/TAZ nuclear fold-change changes more dramatically by downregulating the activation rates of FAK and RhoA than by upregulating them. It also shows that changing FAK activation has a larger impact on YAP/TAZ fold change than changing RhoA activation. The reduction of YAP/TAZ stiffness sensitivity resulting from downregulated FAK (reducing FAK activation rate) can be rescued by upregulating RhoA (increasing the RhoA activation rate). However, the reduction of YAP/TAZ stiffness sensitivity by reducing either the mDia or ROCK activation rate cannot be rescued by increasing the other (Figure S4 in the Supporting Materials). This is due to the fact that mDia and ROCK affect actin and myosin separately. Hence, the model gives rise to a new approach to compare the sensitivities quantitatively.

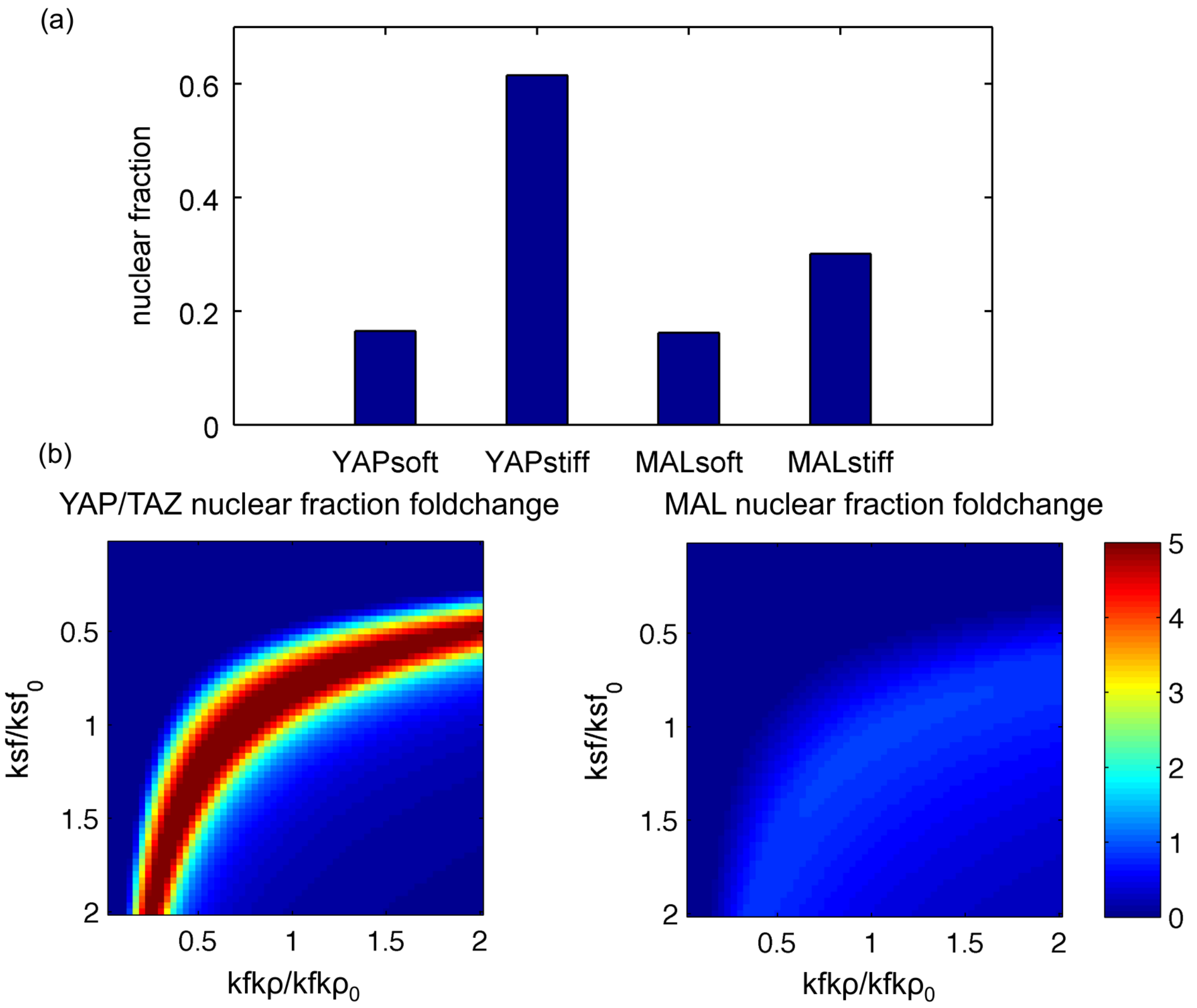

**Figure 3 YAP/TAZ activity sensitivity analysis and comparison with SRF/MAL**

(a) shows YAP/TAZ nuclear localization under different stiffness environment, in comparison with MAL nuclear translocation. Here MAL nuclear translocation is not as robust as YAP/TAZ in response to stiffness change, likely due to YAP/TAZ having a myosin-dependent amplification effect other than the common regulator G-actin/F-actin in SRF/MAL activity. (b) shows fold-change of YAP/TAZ and MAL activities depending on the kinetic parameters, such as the activation rate of FAK ($k_{sf}/k_{sf0}$) and RhoA ($k_{fk\rho}/k_{fk\rho 0}$), when switching matrix from a soft to a stiff one. In addition, YAP/TAZ keeps showing higher sensitivity than MAL translocation fold-change in response to different stiffnesses, and the shape of the activated region reveals that downregulating FAK and RhoA has a more robust effect in regulating the YAP/TAZ nuclear localization than upregulating it.

*The synergistic effect between the Hippo and mechano-sensing pathways*

Our previous analyses are carried out with a constant LATS condition. There is a puzzling fact arising from simultaneous changes in LATS activity and cell tension due to mechanical signals: Dupont et al (8) showed that LATS kinase activity is independent of cell tension, while Aragona et al (7) showed inhibiting LATS and capping proteins for F-actin together can promote YAP/TAZ activity to a greater degree than the total combined effects of adding only siCapZ and adding only siLATS. This is thus a synergy in inhibiting capping proteins and LATS. In the previous work, this phenomenon is characterized as LATS regulation dominated by cytoskeleton regulation (7), however, its mechanism is unknown. Our model predicts that this synergistic effect is from the direct interaction between LATS and LIMK-cofilin. In single-cell system within a soft environment, though most LATS is inactive, LATS can dramatically reduce the phosphorylation of cofilin and enhances actin polymerization in an LATS kinase activity-independent way (64). By adding this interaction (the red interaction 12 in Figure 1), our model is able to capture the synergistic effect between the Hippo and mechano-sensing pathways. Through this interaction, knockout of LATS enhances F-actin assembly. More importantly, it gives rise to the synergistic effect of LATS and CapZ inhibition (Figure 4a). If YAP/TAZ is more dependent on cytoplasmic F-actin than myosin (n changes from 1 to 2 in Eq. 9), the synergistic effect is larger (Figure 4b).

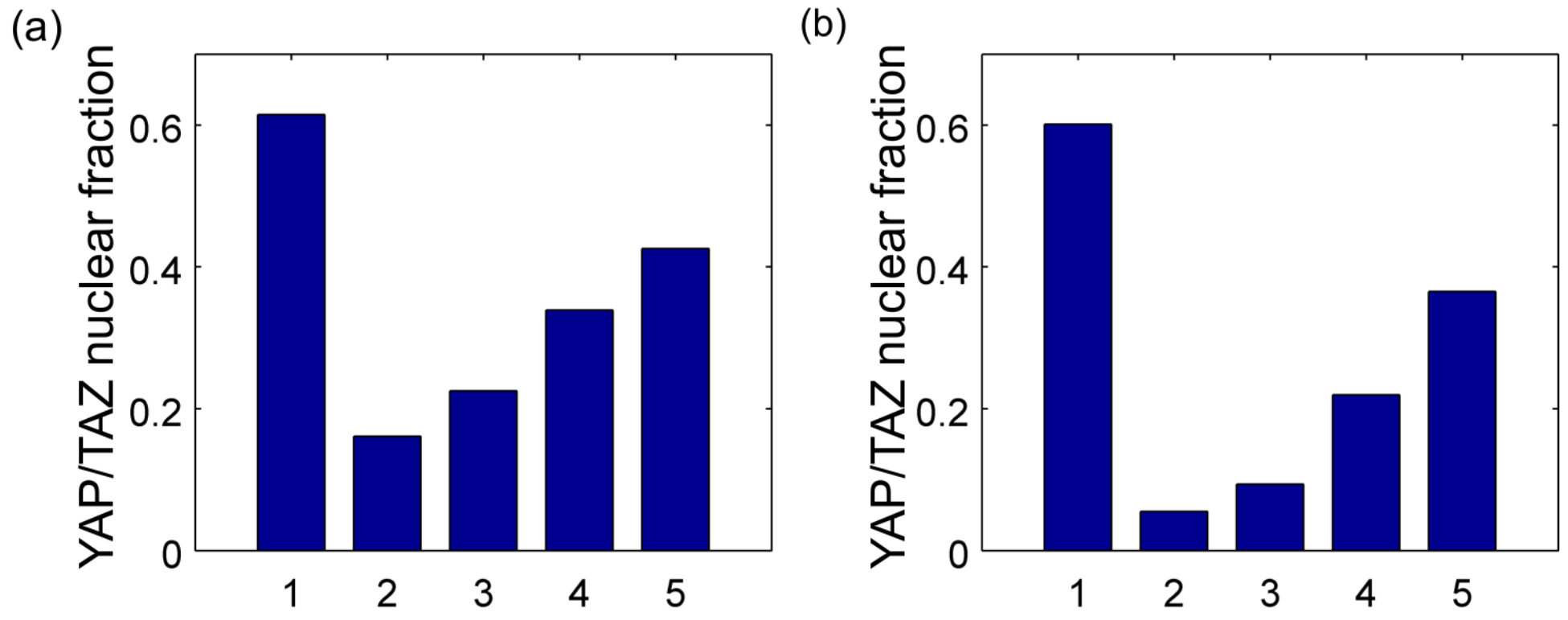

**Figure 4 The synergistic effect between the mechano-sensing and the Hippo pathway.**
(a) Adding the interaction between LIMK and LATS, which induces more free active LIMK molecules by adding siLats, YAP/TAZ is activated synergistically compared with inducing F-actin assembly and adding siLats separately. Condition1: stiff environment. 2: Soft environment. 3: Soft environment with siLats. 4: Soft environment with siCapZ. 5: Soft environment with both siLats and siCapZ. CapZ is a capping protein that inhibits F-actin polymerization. Here the effect of siCapZ on soft environment can be quantified as its corresponding YAP/TAZ activity (H(siCapZ)) abstracting the activity of YAP/TAZ in soft environment (control: H(soft)). The synergistic effect is quantified as H(siCapZ) – H(soft) + H(siLats) – H(soft) > H(siCapZ+siLats) – H(soft), i.e., H(siCapZ) + H(siLats) > H(soft) + H(siCapZ+siLats). (b) With a higher dependence on F-actin than on myosin, the YAP/TAZ showed a more apparent synergistic effect.

# Discussion

The key goal of our study is to understand YAP/TAZ activity as a function of mechanical ECM properties. Here we present a novel mechano-biochemical model that predicts YAP/TAZ activity by integrating adhesion, RhoA GTPases and actomyosin dynamics. The transduction of the mechanical ECM properties into intracellular signaling is quantitatively characterized, and we predict how regulating the signaling molecules can make cells feel different mechanical environments. This gives rise to a direct quantitative function of the mechanical and intracellular molecule signals, the stiffness response function. We have focused on a few core regulators of cytoskeleton dynamics, which are sufficient to predict the difference between cytoskeleton-dependent YAP/TAZ and SRF/MAL activities. Furthermore, the synergistic effect between the LATS and F-actin regulator LIMK can be fully captured by this model. The model indicates that F-actin may have a larger impact in modulating YAP activity than myosin. The bulk of the previous studies on cell-matrix interactions have relied on perturbing signaling molecules in different ECM environment, but mostly characterize only relative upregulating or downregulating effects. Our study thus provides us with methods to systematically study the sensitivity of signaling molecules and the environment with the stiffness response function and fold-change function of different effectors (YAP and SRF). Most importantly, this work provides a quantitative framework for studying YAP/TAZ activity in response to mechanical sensing and predicts consistent results with the synergistic effect of mechanical and canonical Hippo signaling.

There are several interesting aspects and predictions of the model that merit further discussion. A key outcome is that the adhesion is a more robust regulator in affecting YAP/TAZ activities than other cytoskeleton regulators (Figure 2b). This may be due to two reasons: one is because it is the most upstream signaling molecule that initiates the signal transmission and others are not as sensitive to the input signals. The other, and perhaps a more important reason may be that for the RhoA actomyosin cascade, mDia and ROCK are acting complementarily on actin polymerization. A molecule in either of the mDia or ROCK cascades is not as influential in affecting YAP/TAZ as upstream molecules such as FAK or RhoA. This observations is consistent with experimental studies focusing on overexpression of mDia (8) where YAP/TAZ activities in both stiff and soft environment are up-regulated but still have a distinction, due to the additional ROCK cascade. Furthermore, manipulating FAK will change the range of stiffness sensing where YAP/TAZ activity rises with increasing stiffness. However, manipulating mDia or other actin regulators will only raise the activity of YAP/TAZ at soft stiffness, but will not change the range of stiffness sensitivity (Figure S3 in the Supporting Materials). Our model also shows that the upregulation of mDia or deletion of cofilin upregulates YAP/TAZ through inducing contractile actin stress fiber, which is consistent with what has been observed previously (65). Secondly, the inhibition of LATS and CapZ analysis in our model indicates that YAP/TAZ may have a higher dependence on F-actin than myosin. It remains unclear whether this is due to stress fiber formation itself has a larger dependence on F-actin, or whether YAP/TAZ is more sensitive to F-actin than myosin. There have been proposed interactions between myosin and actin cytoskeleton, while whether myosin leads to F-actin assembly (66, 67) or disassembly (34) is not well studied. Our model in its current form does not include any direct interactions between myosin and F-actin. The third interesting prediction is that LATS regulates the F-actin formation via interaction with LIMK/Cofilin, which gives rise to the explanation for the synergistic effect of YAP/TAZ activity between the mechanosensing and the Hippo pathways. This is consistent with some recent findings in *Drosophila*. The overexpression of *hpo* reduces F-actin at apical sites independent of Yki/YAP, and partially suppresses the F-actin accumulation caused by knocking down the capping proteins (57). By inhibiting the binding of LIMK and LATS while preserving the integrity of their individual phosphorylation function, the synergistic effect of the Hippo and the mechanical transduction will be greatly reduced or diminished. Other actin regulators, such as ENA (68) and zyxin (69), were found to affect Yki/YAP activity synergistically (70), most likely through their actin bundles and stress fiber regulation as well. The experimental study of the YAP/TAZ regulation cascade is difficult, due to the existence of the co-existing alternative interactions and pathways mentioned above. With our computational model, we are able to investigate the proposed interactions or possible mechanisms for the YAP/TAZ regulation in a more comprehensive way. This method facilitates an understanding of the complicated interactions and pathways in an alternative and efficient manner.

There are also some parts in need of further clarifications. Currently the model mostly studies the single cell stiffness-sensing mechanism in an EGF-sufficient environment, and is not focused on multicellular environments or cell-cell adhesion or stress. The changing of EGF or serum dose can also be rate-limiting factors for this stiffness sensing and YAP/TAZ signaling pathway and is not considered in the model. That said, as mentioned in the methods section, our model can be extended to include other molecules, such as the Hippo upstream regulators or PI3K, which are also known to affect YAP/TAZ activity. Secondly, it has been demonstrated that expressing kinase-dead FAK in FAK-/- cells prevented growth, which is consistent with our model. However, FAK-/- cells exhibit an excessive adhesion formation and uncontrolled proliferation (52). This suggests that the absence of FAK, rather than inactive FAK, triggers an alternative pathway activation (51, 52, 71). Nevertheless, the FAK-null cells have been shown to be unable to detect soft versus rigid substrata (72). Thus here our predictions and analysis of FAK activity are restricted to the regime of the downregulation of its total amount or the phosphorylation rate change, rather than the complete knockout of FAK. There have also been studies reported in literature showing the mechanical feedback of F-actin stiffening on ECM properties (73). However, the role of biochemical signaling transduction during this process is not well known and thus this feedback is not currently included in the model.

Overall, our model provides a platform to study mechanical regulation and molecular crosstalk of YAP/TAZ upstream regulators. It will be interesting to combine our stiffness-sensing model with models such as the effect of growth factor stimulation to explore potential interactions between various stimuli. Likewise, it is also important to extend the model in a multicellular context to capture the *in vivo* complexity. To this purpose, incorporating cell-cell adhesion and its effect on the dynamics of the canonical Hippo pathway into our model will provide useful insights. Considering that the mechanical effects can regulate YAP/TAZ independent of the canonical Hippo pathway, and the cell size control mechanism is very similar to the one of stiffness sensing (7), we believe that our current model can be useful and quantitatively predictive in a multicellular environment in elucidating key processes that are too costly to investigate or remain experimentally intractable.

# Author Contributions

M.S., F.S. and M.H.Z developed the concepts and designed the models, M.S. implemented the simulations, M. S. and F. S. performed data analysis, and M.S, F.S. and M.H.Z wrote the article.

# Acknowledgements

The authors acknowledge research support from the National Institutes of Health (U01-CA177799) and the National Science Foundation (DMR-1206635) for this work. We deeply appreciate the input from members of our lab during this research, and especially Robert J. Seager and Dan Reynolds for useful comments on our manuscript.

# Supporting Citations

References (74–80) appear in the Supporting Material.

# Supporting Materials

## Supporting Methods

Molecular intervention

When latrunculin A is employed *in vitro*, the depolymerization of actin filaments results from sequestration of monomeric actin by latrunculin (1), and the effect in our model is mimicked by changing the polymerization rate of F-actin in the cytoplasm to 1/10 of the original. Overexpression of mDia is mimicked by changing $mDia_0$ to 5-fold. Blebbistatin and Y-27632 inhibit myosin and ROCK, thus we mimic their effects by changing the deactivation rates of myosin $k_{dmy}$ and the one of ROCK to 10-fold.

Local sensitivity analysis of YAP/TAZ activity on kinetic parameters

Each kinetic parameter is locally up- or down-regulated by 10% individually, the YAP/TAZ activity, quantified by YAP/TAZ nuclear translocation fraction, is calculated

$$ParameterSensitivity = \frac{\left|YAP/TAZ_N(k+\Delta k) - YAP/TAZ_N(k)\right|}{YAP/TAZ_N(k)} / (\frac{\Delta k}{k}).$$

Here $\Delta k / k = 10\%$. If parameter sensitivity is 1, the dependence on the correspondent kinetic parameters is linear. If it is larger than 1, it shows the YAP/TAZ activity is highly dependent on the parameters in a way that is more dramatic than the linear dependence, while if it is smaller than 1, then it is the other way around (2).

## Supporting Figures

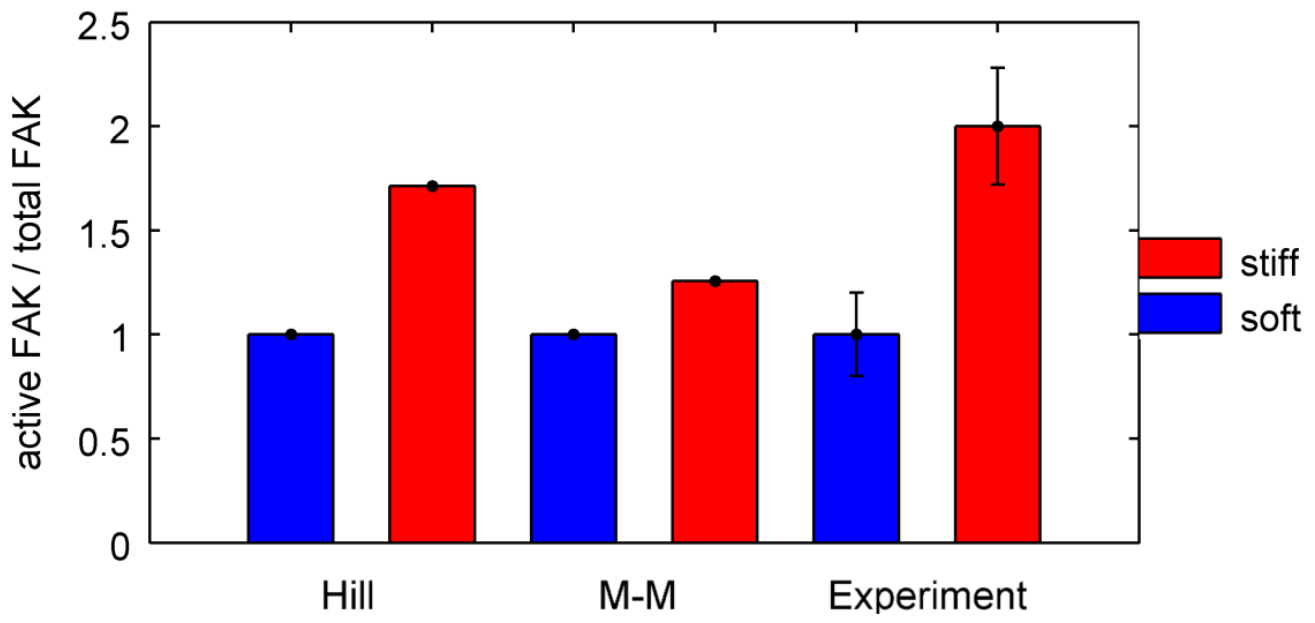

**Figure S1 The FAK activity model comparison.**

Our model adapts from the previous Michaelis-Menten model of FAK activation rate dependent on ECM ligand density (3, 4). After incorporating the ECM stiffness in the model, we found that the second-order hill function matches the experimental data from previous work (5), with the Michaelis-Menten model not able to capture the increase of FAK activity from soft to stiff environments.

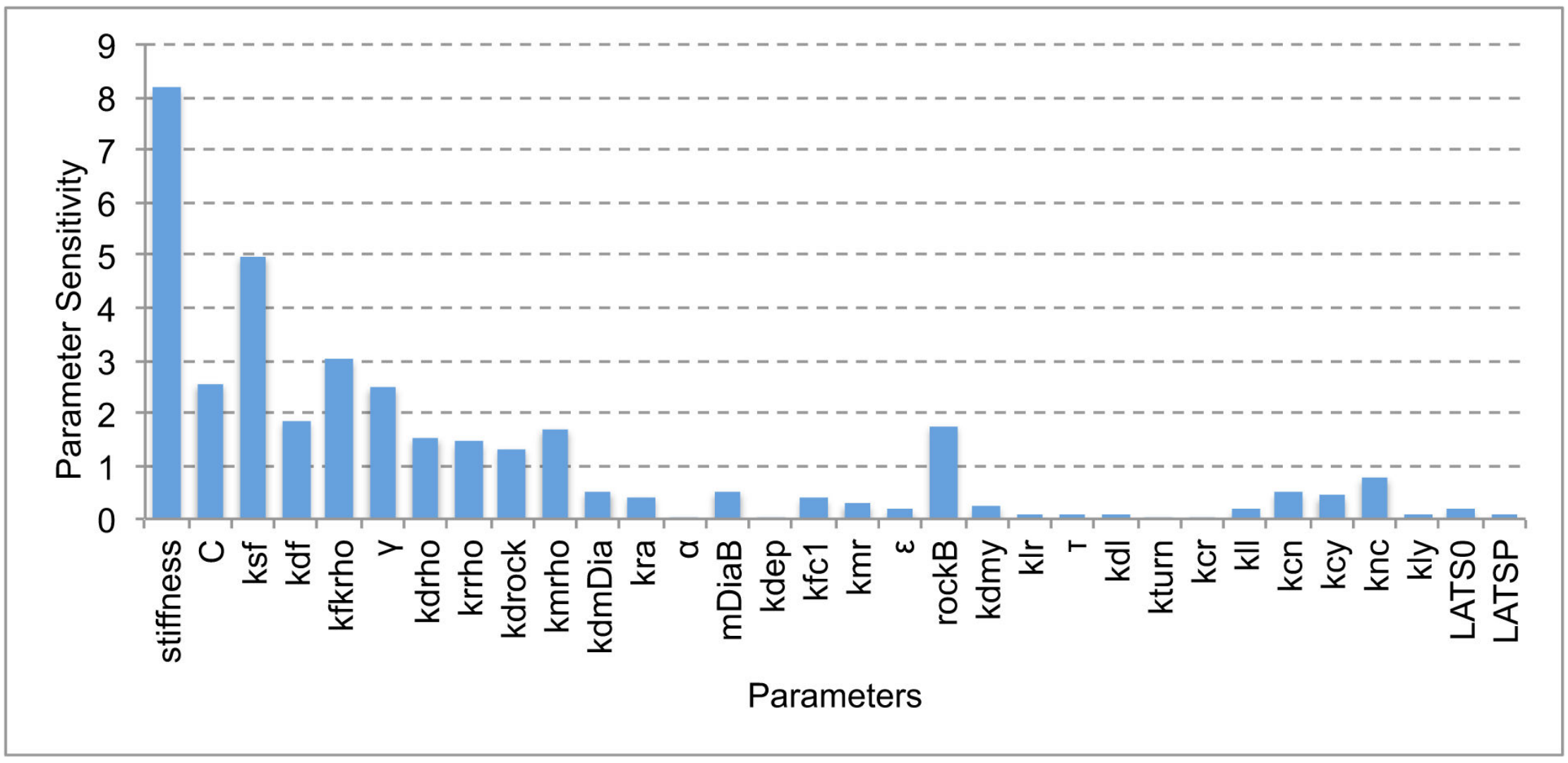

**Figure S2 Local sensitivity analysis of YAP/TAZ model in soft environment**

This shows the local sensitivity in YAP/TAZ model by up-regulating the kinetic parameters in a soft environment. In consistent with Figure 2b, the adhesion components in the model have the largest impact on the YAP/TAZ activity.

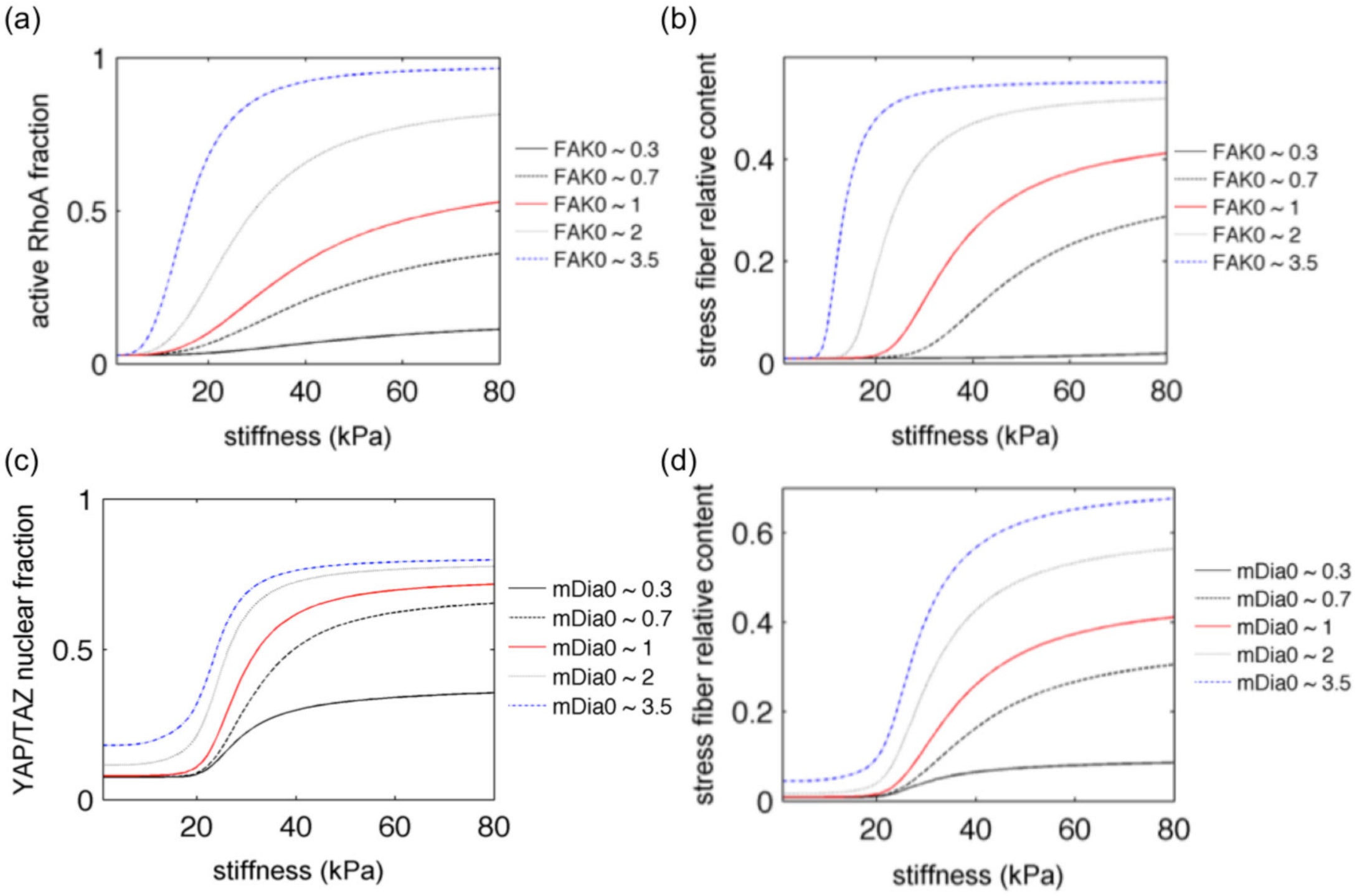

**Figure S3 stiffness response function with changing protein contents**

(a) and (b) are active RhoA fraction and stress fiber content changing along with the stiffness under different amount of FAK. The RhoA activity goes plateau as the stiffness increases. Both the sensitivity range and the plateau position of RhoA change with FAK intervention. (c) The YAP/TAZ stiffness response function shifts vertically by changing total amount of mDia for 3D NMuMG cells as an example. mDia0~1 corresponds the endogenous mDia amount of NMuMG cell in 3D collagen. (d) The stress fiber content stiffness response function changes along the total amount of mDia. The stress fiber content change with stiffness gives rise to the YAP/TAZ stiffness response. As predicted in YAP/TAZ activity, the FAK shifts the stress fiber content stiffness response function horizontally, while other actin molecules such as mDia shifts the stress fiber content stiffness response function vertically.

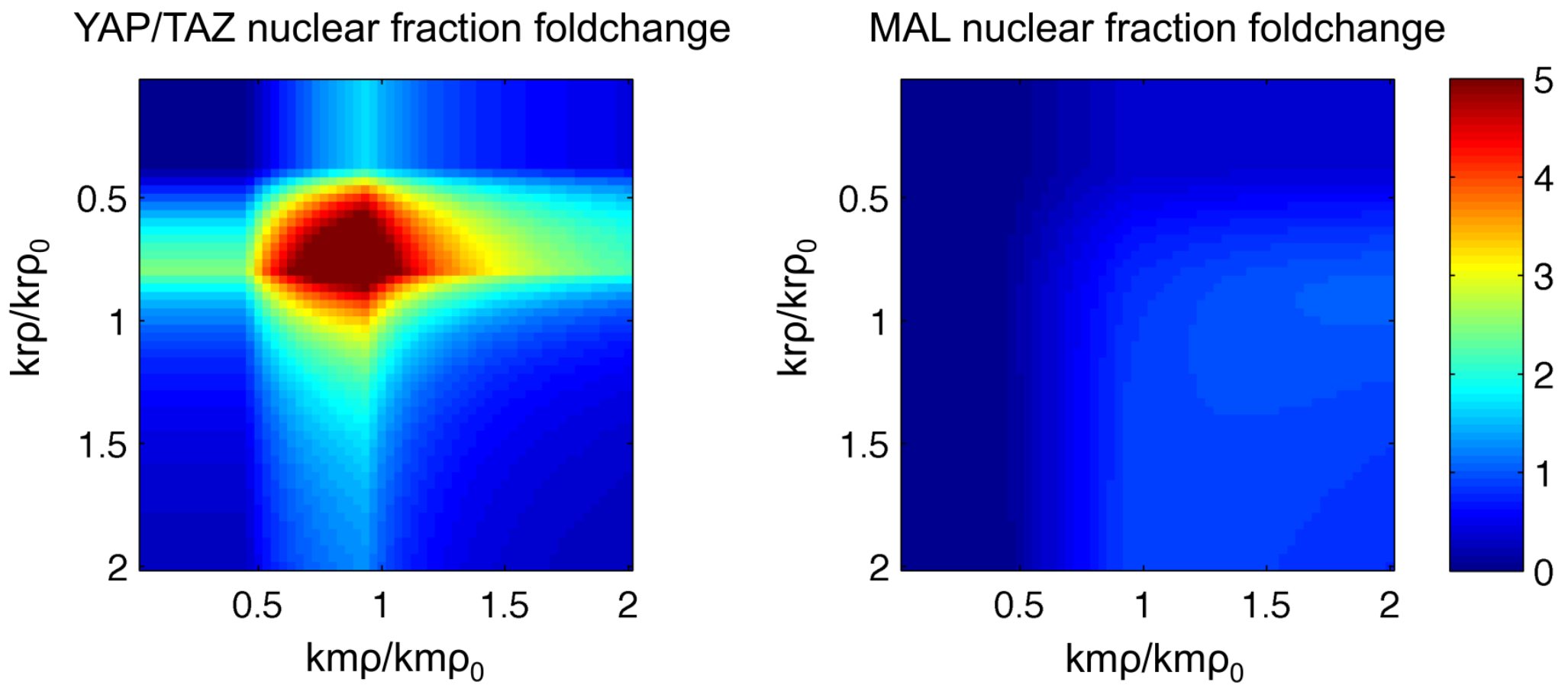

**Figure S4 The kinetic parameter analysis of ROCK and mDia**

Using the same method as in Figure 3b, we analyze the activation rate of ROCK and mDia on YAP/TAZ sensitivity to ECM stiffness, and also compare with SRF/MAL sensitivity. Similar with regulating adhesion and RhoA components, down-regulating ROCK and mDia shows a dramatic effect in the YAP/TAZ sensitivity to stiffness. However, SRF/MAL shows a relative monotonic dependence on mDia activation, which regulates F-actin polymerization. This indicates that SRF/MAL increase rate dependence on F-actin is also monotonic and the increase rate of SRF/MAL in stiff environment is larger than the one in soft environment. The YAP/TAZ sensitivity to stiffness pattern shows that by down-regulating the ROCK and mDia, the YAP/TAZ decrease rate in soft environment is first lower than the one in stiff environment, and then rises the other way. Comparing with the dependence on FAK and RhoA sensitivity analysis, the region of high YAP/TAZ activity is smaller, indicating that there are other more robust regulators limiting the active regime of YAP/TAZ activity.

## Supporting nondimentionalized model

The nondimentionalization has been carried out for the model. The refined parameters are in the Table S1. And we mainly focus on the ECM stiffness in our paper and thus treat ligand density constant. The model is applicable with changing ligand density, as it is adapted from the ligand density dependent adhesion model (3, 4). The variables of proteins are normalized per cell. Most of the proteins involved in YAP/TAZ stiffness sensing have been found to be constant under stiff and soft environment, for example, the total amount of FAK (2), RhoA (2), total actin (3), and YAP/TAZ (4).

$$\frac{dfak(t)}{dt} = k_{sf}\frac{E_{mol}(t)^2}{\bar{C}^2 + E_{mol}(t)^2}(1 - fak(t)) - k_{df}fak(t) \qquad \text{(S1)},$$

$$\frac{drhoA(t)}{dt} = k_{fk\rho}(\overline{\gamma}(fak(t))^2 + 1)(1 - rhoA(t)) - k_{d\rho} rhoA(t) \quad \text{(S2)},$$

$$\frac{drock(t)}{dt} = \overline{k_{r\rho}} rhoA(t)(1 - rock(t)) - k_{drock} rock(t) \quad \text{(S3)},$$

$$\frac{drock(t)}{dt} = \overline{k_{r\rho}} rhoA(t)(1 - rock(t)) - k_{drock} rock(t) \quad \text{(S4)},$$

$$\frac{dmyo(t)}{dt} = k_{mr}(\overline{\varepsilon} T(rock(t)) + 1)(1 - myo(t)) - k_{dmy} myo(t) \quad \text{(S5)},$$

$$\frac{dlimk(t)}{dt} = k_{lr}(\overline{\tau} T(rock(t)) + 1)(1 - limk(t)) - k_{dl} limk(t) \quad \text{(S6)},$$

$$\frac{dcofilin(t)}{dt} = k_{turn-over}(1 - cofilin(t)) - \overline{k_{cr}}(1 - \overline{k_{ll}} lats_0) limk^2(t) cofilin(t) \quad \text{(S7)},$$

$$\frac{dfcyto(t)}{dt} = k_{ra}(\overline{\alpha} T(mdia(t)) + 1)(1 - factin(t)) - (k_{dep} + \overline{k_{fc1}} cofilin(t)) fcyto(t) \quad \text{(S8)},$$

$$\frac{dyaptaz(t)}{dt} = (k_{CN} + \overline{k_{CY}} fcyto^n(t) myo_A(t))(1 - yaptaz(t)) - (k_{NC} + \overline{k_{ly}} lats(t)) yaptaz(t)$$

(S9),

$$\frac{dmal(t)}{dt} = \frac{\overline{k_{cnm}}}{(1 - factin(t))^2}(1 - mal(t)) - k_{ncm} mal(t) \quad \text{(S10)},$$

**Table S1. The parameter values used in the nondimentionalized model.**

| Parameter | Value | Reference |
|---|---|---|
| $k_{sf}$ | 0.0175 $s^{-1}$ | Kumar et al (4) |
| $k_{df}$ | 0.035 $s^{-1}$ | Kumar et al (4) |
| $\overline{C}$ | 45 kPa for 3D NMuMG cells | Estimated from Provenzano et al (5) |
| $\overline{C}$ | 25 kPa for 2D MEC and MSC | Estimated from Dupont (6) |
| $k_{fk\rho}$ | 0.018 $s^{-1}$ | Estimated from Provenzano et al (5) |
| $\overline{\gamma}$ | 500 | Estimated from Provenzano et al (5) |
| $k_{d\rho}$ | 0.625 $s^{-1}$ | Sako et al (7) |
| $\overline{k_{r\rho}}$ | 2.2 $s^{-1}$ | Edelstein-Keshet et al (8, 9) |
| $k_{drock}$ | 0.8 $s^{-1}$ | Estimated from Feng et al (10) and Ji et al (11) |
| $\overline{k_{m\rho}}$ | 1 $s^{-1}$ | Estimated |
| $k_{dmDia}$ | 1 $s^{-1}$ | Estimated |
| $k_{mr}$ | 0.015 $s^{-1}$ | Estimated |
| $\bar{\varepsilon}$ | 40 | Estimated |
| $k_{dmy}$ | 0.067$s^{-1}$ | Cirit et al (12) |
| $k_{lr}$ | 0.07 $s^{-1}$ | Estimated from Tania et al and Song et al (13, 14) |
| $\bar{\tau}$ | 200 | Estimated |
| $k_{dl}$ | 2 $s^{-1}$ | Estimated from Tania et al and Song et al (13, 14) |
| $k_{turn-over}$ | 0.04 $s^{-1}$ | Tania et al (13) |
| $\overline{k_{cr}}$ | 0.7 $s^{-1}$ | Tania et al (13) |
| $\overline{k_{ll}}$ | 0.8 $s^{-1}$ | Estimated from Yang et al (15) |
| $k_{ra}$ | 0.4 $s^{-1}$ | Estimated |
| $\bar{\alpha}$ | 40 | Estimated from Higashida et al (16) |
| $k_{dep}$ | 3.5 $s^{-1}$ | Bonder et al (17) |
| $\overline{k_{fc1}}$ | 8 | Estimated from Tania et al and Song et al (13, 14) |
| $k_{CN}$ | 0.1 $s^{-1}$ | Estimated |
| $\overline{k_{CY}}$ | 20 $s^{-1}$ | Estimated |
| $k_{NC}$ | 3 $s^{-1}$ | Estimated |
| $\overline{k_{ly}}$ | 6 $s^{-1}$ | Estimated |
| $\overline{k_{cnm}}$ | 0.2$s^{-1}$ | Estimated from Connelly et al (18) |

**Table S2. Redefinition of the parameters in the nondimentionalized model**

| Parameters in the nondiment-ionalized model | Parameters in the dimentionalized model | Parameters in the nondiment-ionalized model | Parameters in the dimentionalized model |
|---|---|---|---|
| $\bar{C}$ | $C/LD$ | $\overline{k_{cr}}$ | $k_{cr}LIMK_0^2$ |
| $\bar{\gamma}$ | $\gamma FAK_0^2$ | $\overline{k_{ll}}$ | $k_{ll}LATS_0$ |
| $\overline{k_{r\rho}}$ | $k_{r\rho}RhoA_0$ | $\bar{\alpha}$ | $\alpha mDia_0$ |
| $\overline{k_{m\rho}}$ | $k_{m\rho}RhoA_0$ | $\overline{k_{fc1}}$ | $k_{fc1}Cofilin_0$ |
| $\bar{\varepsilon}$ | $\varepsilon ROCK_0$ | $\overline{k_{CY}}$ | $k_{CY}actin_0^n Myo_0$ |
| $\bar{\tau}$ | $\tau ROCK_0$ | $\overline{k_{ly}}$ | $k_{ly}LATS_0$ |
| $\overline{k_{ra}}$ | $k_{ra}(1+\beta)$ | $\overline{k_{cnm}}$ | $k_{cnm}/k_{mg}(1+\beta)^2$ |
| $fak$ | $FAK_P / FAK_0$ | $limk$ | $LIMK_A / LIMK_0$ |
| $rhoA$ | $RhoA_{GTP} / RhoA_0$ | $cofilin$ | $Cofilin_{NP} / Cofilin_0$ |
| $rock$ | $ROCK_A / ROCK_0$ | $fcyto$ | $Fcyto / actin_0$ |
| $mdia$ | $mDia_A / mDia_0$ | $yaptaz$ | $(YAP/TAZ)_N / (YAP/TAZ)_0$ |
| $myo$ | $Myo_A / Myo_0$ | $mal$ | $MAL_N / MAL_0$ |

**Supporting Reference**